# Designing an Adaptive Storytelling Platform to Promote Civic Education in Politically Polarized Learning Environments

Christopher M. Wegemer[1], Edward Halim[1], and Jeff Burke[1]

[1] University of California, Los Angeles, Los Angeles, CA 90095, USA
cwegemer@ucla.edu

**Abstract.** Political polarization undermines democratic civic education by exacerbating identity-based resistance to opposing viewpoints. Emerging AI technologies offer new opportunities to advance interventions that reduce polarization and promote political open-mindedness. We examined novel design strategies that leverage adaptive and emotionally-responsive civic narratives that may sustain students' emotional engagement in stories, and in turn, promote perspective-taking toward members of political out-groups. Drawing on theories from political psychology and narratology, we investigate how affective computing techniques can support three storytelling mechanisms: transportation into a story world, identification with characters, and interaction with the storyteller. Using a design-based research (DBR) approach, we iteratively developed and refined an AI-mediated Digital Civic Storytelling (AI-DCS) platform. Our prototype integrates facial emotion recognition and attention tracking to assess users' affective and attentional states in real time. Narrative content is organized around pre-structured story outlines, with beat-by-beat language adaptation implemented via GPT-4, personalizing linguistic tone to sustain students' emotional engagement in stories that center political perspectives different from their own. Our work offers a foundation for AI-supported, emotionally-sensitive strategies that address affective polarization while preserving learner autonomy. We conclude with implications for civic education interventions, algorithmic literacy, and HCI challenges associated with AI dialogue management and affect-adaptive learning environments.

## 1 Introduction

Civic learning has become impeded by rising political polarization [1, 2], exacerbated by self-selection of digital content and misinformation [3]. Large Language Models (LLMs) have demonstrated potential for personalizing civic content in a way that may reduce polarization [4], yet most interventions rely on rational argumentation, despite leading evidence suggesting that polarization is primarily driven by identity-based animosity [5, 6]. Adaptive digital storytelling may uniquely support students' democratic open-mindedness and reduce polarization by promoting transportation into a narrative, identification with a character who holds differing perspectives, and interaction with the storyteller.

We explore how techniques from affective computing [7, 8] can be leveraged to develop a novel civic storytelling platform for secondary and post-secondary students. Synthesizing theories of narrative persuasion [9, 10] with design strategies from adaptive storytelling [11, 12, 13], our AI-mediated Digital Civic Storytelling (AI-DCS) prototype combines computer vision, LLMs, and machine learning to dynamically respond to users' emotional and attentional states, supporting personalized and emotionally-responsive podcast-style civic stories. In contrast to storytelling research that prioritizes adaptive plot sequences [14], AI-DCS centers beat-by-beat adjustments in narrative language to modulate users' emotional engagement. Our AI-DCS platform addressed three design tensions inherent to civic education in polarized contexts: (1) young people's engagement with diverse political perspectives is constrained by their own political identity; (2) emotional engagement is necessary to promote open-mindedness, yet eliciting emotions risks exacerbating polarized tensions; (3) current interventions based on intergroup dialogue require skilled facilitation and are limited to small groups, which constrains scalability. These tensions are briefly presented below, alongside theories that provide a conceptual foundation for our approach. The remainder of our paper describes the implementation and testing of our prototype architecture.

# 2 Theoretical Background

## 2.1 Tensions of Political Polarization in Civic Education

Civic education is vital for preparing youth to participate in democratic society [15]. Across civic education frameworks, the ability of young people to critically engage with diverse political perspectives is a central goal [16, 17]. However, acrimonious political tensions in educational environments have increasingly impeded civic learning and undermined the civic mission of schools [2].

The growing political divide in the US is primarily attributable to affective, rather than cognitive, polarization [5, 6]. Drawing from social identity theory [18], affective political polarization is defined as animosity towards out-group partisans and affiliation with in-group partisans [6]. In polarized educational settings, civic interventions that challenge learners' pre-existing beliefs can trigger defensive reactions [19], and even subtle political cues can lead students to be dismissive of instruction [20, 21]. Young people tend to filter political information in relation to how strongly it aligns with their worldview and group membership [22]. The dependence of students' reception of civic content on their own polarized identities poses a fundamental challenge for promoting pluralistic tolerance and open-mindedness [1, 23].

Scholars of civic education have increasingly recognized that democratic attitudes require socioemotional skills [24] and that rational argumentation has limited efficacy [25]. Fostering emotional engagement in civic content can serve as a catalyst for learning that promotes openness to new ideas [26, 27, 28]. However, centering sensitive and controversial issues in a classroom setting risks triggering overwhelming and conflicting emotional responses that exacerbate partisan contentions and overwhelm cognitive processing [29].

Current approaches to addressing polarization predominantly rely on scaffolding interactions between students with diverging political beliefs. Intergroup contact theory suggests that positive interactions between members of opposing political groups can foster understanding of others' lived experiences, which reduces animosity [30, 31]. However, without appropriate structure, intergroup dialogue risks backfiring and exacerbating tensions [29]. The technique is inherently limited to small groups of students because the interactions are challenging to facilitate, resource-intensive, and conditional on students' willingness to engage with opposing points of view [29, 32]. The emergence of responsive AI that mimics human emotional responsiveness presents new opportunities and design questions for educational interventions. Specifically, AI-mediated storytelling represents a scalable approach that may cultivate personalized and predictable emotional engagement in civic content.

### 2.2 The Potential of Storytelling to Foster Emotional Engagement

Emotional engagement in stories has been linked to long-term changes in political attitudes and behaviors [9, 33, 34], even on controversial issues [10, 35, 36]. Informed by research on narratology and narrative persuasion, our AI-DCS approach leverages three storytelling mechanisms to foster students' emotional engagement in civic stories.

First, the persuasive power of a story hinges in part on the extent to which listeners experience ***transportation*** into the narrative [37, 38]. When listeners are deeply immersed in a story, the concurrent suspension of reality reduces the likelihood of reactionary counterarguing [9] and supports engagement with unfamiliar perspectives without making their political identities vulnerable [39]. Subtle shifts in descriptive and emotive language can enhance transportation [40, 41], and such linguistic changes can be adjusted in response to listeners' real-time affective states.

Second, listeners' ***identification*** with story characters has been shown to facilitate attitudinal change [42, 43]. Even when a character holds differing political beliefs, listeners who identify with them may vicariously experience their struggles and growth, which enhances empathy and cross-partisan understanding [33, 44]. Listeners are more likely to identify with characters when they perceive shared characteristics [45]. To facilitate identification, AI-generated character voices can be tailored to reflect demographic characteristics and personal preferences of the listener (i.e., perceived importance of specific social issues).

Third, ***interaction*** with the storyteller can deepen listeners' engagement with the narrative [46]. Research on human-AI interaction suggests that students readily form parasocial bonds with emotionally attuned AI agents [47, 48] and display affective engagement comparable to peer or teacher conversations [49]. In our approach, an AI narrator delivers the story via conversational audio exchange and behaves as a responsive peer who poses reflective questions, reacts intuitively to listeners' emotions, and adjusts tone accordingly.

Our design conjectures incorporate these three storytelling mechanisms, for which AI tools may be effectively applied to facilitate students' emotional engagement in political narratives (see Table 1). Additionally, we used well-established storytelling conventions that promote emotional engagement, such as dramatic arc structure [50] and first-person perspective [51]. Current technology supports dynamic adaptation and personalization of voice-based narratives [52]. Further, podcast-style storytelling enables educators to engage students emotionally with minimal preparation or risk of unintended conflict [53].

**Table 1.** Proposed story adaptation mechanisms.

| Story features | Data used | Adaptations | Timing of adaptation |
|---|---|---|---|
| Emotive and descriptive language | Facial emotion recognition | Story language is adjusted to facilitate ***transportation*** into the narrative if a mismatch between expected and actual emotional reactions of student is detected | Dynamically during story |
| Narrator (Main character) | Identity characteristics ascertained from dialogue | Narrator characteristics match the student's to promote ***identification*** with the character | Prior to start of story |
| Dialogue with narrator | Supervising of student dialogue | Through conversational ***interaction***, the narrator asks student about their experience of the story if student is persistently inattentive or emotionally disengaged | Intermittently as needed to re-engage the student |

### 2.3 Advancing Affective Computing and Adaptive Storytelling

The present work draws from advances in affective computing to operationalize emotional engagement as a dynamic input into civic learning design. Affective computing enables systems to sense, interpret, and respond to users' emotional states in real time [54, 55]. Learning systems that respond adaptively to learners' affective signals have shown promise for improving engagement, persistence, and learning gains across a range of domains [56, 57]. However, most educational applications of affective computing have focused on STEM and language learning. Civic education introduces qualitatively different affective dimensions and design constraints. Rather than optimizing for task performance or reducing frustration in cognitive tasks, we leverage storytelling to modulate emotional engagement in ways that support perspective-taking while preventing affective disengagement or identity-protective resistance. That is, we aim to identify how affective computing can be leveraged not only to scaffold learning, but to navigate complex emotional and identity-based barriers to civic dialogue.

Our AI-DCS prototype extends affective computing into the civic domain by integrating facial emotion recognition to monitor learners' affective responses to politically charged narratives. Utilizing computer vision, current emotion detection models have demonstrated high accuracy at classifying users' explicit and implicit emotional states in real-time [7, 58, 59]. Convolutional neural networks (CNNs) have demonstrated accuracy above 99% in classifying facial emotions in real-world contexts [60] and have also been incorporated in most recent mixed-media foundation models, i.e., GPT-4o.

Similarly, attention detection models have demonstrated high accuracy in educational platforms.

Our work is located at the intersection of affective computing and adaptive storytelling. Adaptive storytelling systems aim to dynamically modify narrative content in response to real-time user states, preferences, or behaviors [11, 12]. Much of the prior work in this domain has focused on plot adaptation, character agency, or branching storylines within entertainment or game-based contexts. In contrast, we focus on beat-by-beat modulation of narrative language to sustain emotional engagement with potentially sensitive political content. The adaptive layer operates at the level of linguistic framing, adjusting specific language in response to users' emotional alignment with expected empathic responses. This fine-grained, localized adaptation seeks to maintain the learner within an optimal affective window, supporting sustained engagement in out-group perspectives without triggering affective overload or defensive disengagement.

Research on technology-assisted dialogue in classrooms suggests that adaptive tools can promote inclusive participation and reduce interpersonal tensions among students [61]. Through the development of our prototype, we seek to answer the question: How can adaptive design features be implemented to facilitate emotional engagement in stories that center experiences of political out-group members? Using novel adaptive storytelling techniques, we aim to provide scaffolding that fosters students' perspective-taking and critical civic empathy [27] by dynamically responding to learners' affective states in response to first-person podcast-style narratives about pressing social issues.

## 3 Method

Situated within UCLA's Center for Research in Engineering, Media and Performance, our team consisted of a postdoctoral researcher (Chris Wegemer), an undergraduate researcher (Edward Halim), and a professor (Jeff Burke). We employed a design-based research (DBR) process [62] to develop and refine an emotionally-responsive civic storytelling platform that educators could utilize with secondary and postsecondary students. DBR is well-suited for the development of complex educational technologies situated in real-world use settings, enabling close coupling of interdisciplinary theory with emerging technical affordances. This paper focuses on the technical development and early-stage design testing, which will support formal user testing with diverse student populations in forthcoming studies. We followed the first two phases of the Integrative Learning Design Framework [63], each described in turn.

First, we conducted an informed exploration to identify tools to support our adaptive storytelling mechanisms. Building on an earlier study of college students' changes in political attitudes in response to AI-generated stories [64], we synthesized research across literatures on civic education, affective computing, and narrative persuasion to identify potential features of an emotionally-adaptive civic storytelling platform (as presented in Table 1). We reviewed existing narrative interventions that aimed to reduce polarization, particularly those that leveraged emotional identification with story characters. Finally, we decided on specific modules to structure the architecture of our adaptive storytelling system.

Second, undergraduate researcher (Edward Halim) constructed the architecture for our AI-DCS platform (see Figures 1, 2, and 3). Through iterative design cycles, Edward tested and refined user interactions with the AI narrator from the perspective of a single developer-researcher. The iterations included running simulated sessions with varied emotional inputs, reviewing system logs and emotion classification outputs, adjusting narrative prompts and emotional thresholds, and modifying rules governing language adaptation and listener re-engagement. The design cycles were grounded in both formative data on system behavior and theoretical expectations of narrative engagement. Our work yielded a prototype that engages users through two stages, initial onboarding and personalized story narration, each described below.

# 4 Platform Architecture and Implementation

## 4.1 Stage 1: Onboarding and Personalization

First, the introductory stage of the prototype (see Figure 1) employs semi-structured verbal interactions with GPT-4 to set dialogic expectations and conversationally ascertain users' political orientation. Based on the user's input, a pre-established story outline ("Story.txt" file) is selected from a repository to facilitate engagement with political out-group perspectives on a social issue that is important to the user. The outlines were derived from narratives co-designed with GPT-3.5 [64]. Similarly, the demographic characteristics of the user can be ascertained to match the main character's demographic and vocal features prior to the start of the story (i.e., age, gender, and race/ethnicity). The onboarding and personalization stage involves the implementation of two key design features, described in turn.

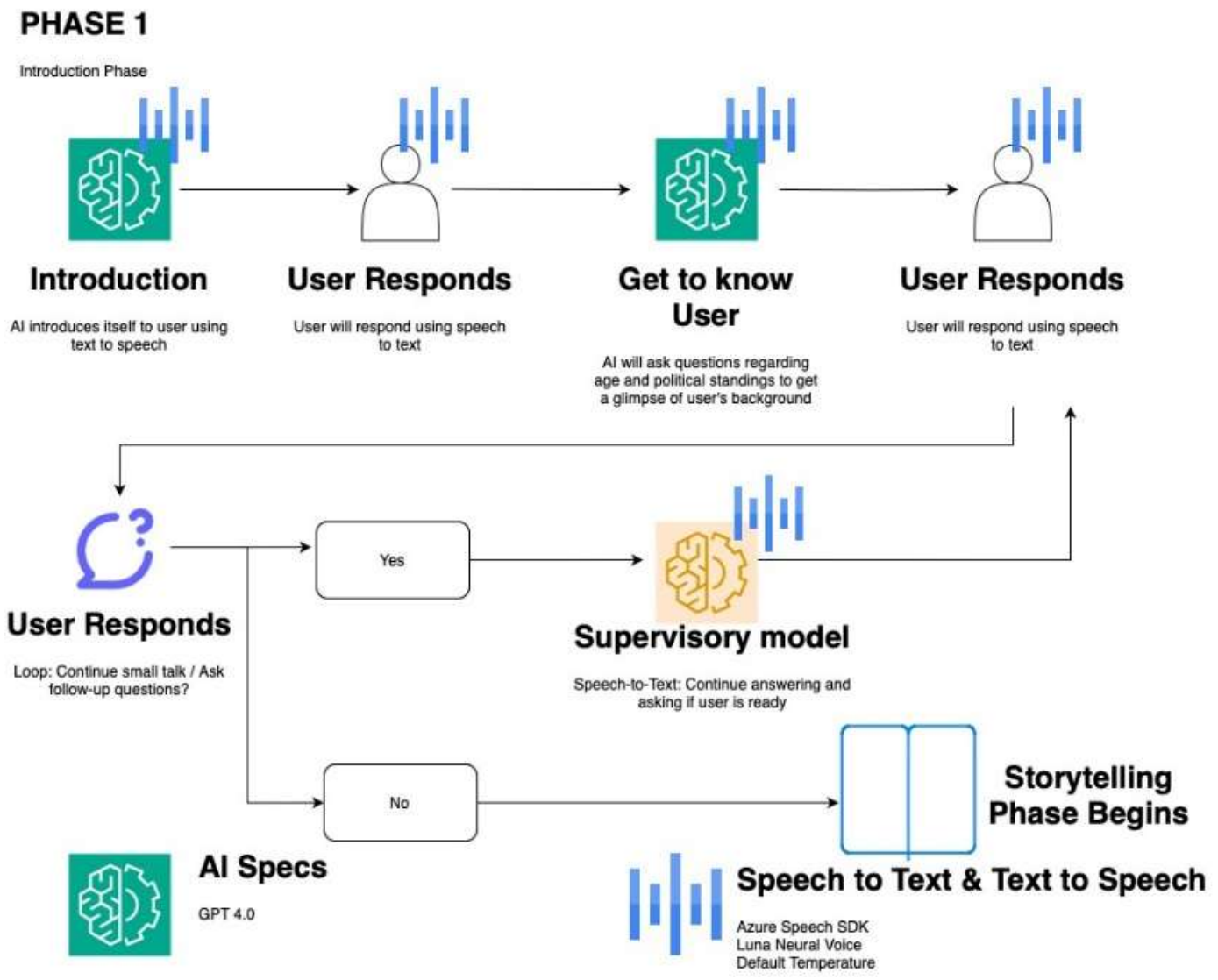

**Fig. 1.** Workflow diagram of the introduction phase of the AI-DCS prototype.

**Feature 1: Human-AI dialogue.** Emotional engagement depends on natural turn-taking responsiveness, and accordingly, the interactivity of the platform was based on conversational exchange. We began developing our prototype before mixed-media foundation model APIs were publicly available and we built the conversational feature by chaining text-to-speech and speech-to-text modules. Specifically, we used Azure Speech SDK to transcribe users' spoken words into text that was passed to GPT-4, then the text output from GPT-4 was converted into speech using Azure Neural TTS. After testing a variety of services, we chose Azure's STT and TTS because they provided accurate transcription and realistic voices with low latency, more so than most other options at the time. We retained our design after dialogic models became publicly available because our approach allowed us to have greater experimental control. Lastly, we used LangChain [65] to simplify GPT prompting as well as record conversation history, which was included in prompts to the AI narrator to provide context for further exchanges. The conversation history also functioned as a qualitative data collection mechanism that will support future user studies.

**Feature 2: Story retrieval.** The onboarding stage ends with the tacit selection of a story that will facilitate the user's encounter with a differing political perspective. Each Story.txt file contains the outline of a narrative and emotional metadata, divided into beat-by-beat segments (typically 2-5 sentences). Each segment is numerically labeled to provide an index for sequential retrieval. The primary emotion that characterizes the segment is labeled, which is retrieved simultaneously and later used to assess whether users' emotional states are consistent with expectations from the story material. (See sample story material in Appendix A.) Using a LangChain approach, each sequential story segment is integrated into a prompt that is sent to GPT-4 for emotionally-responsive personalization, then vocalized for the user.

### 4.2 Stage 2: Emotionally-Adaptive Storytelling

In the storytelling stage (see Figure 2), the AI narrator vocalizes the narrative using a first-person perspective. The user's emotional state and attention are continuously monitored via facial emotion recognition. Emotional states are assessed every 0.1s, averaged over each story segment, compared to expected reactions. If emotional alignment and attention are maintained within a predefined tolerance threshold and the user does not interrupt, the narration proceeds segment-by-segment from the Story.txt outline. Emotional mismatch or lack of attention triggers a prompt that adjusts the language of the subsequent story segment to enhance emotionally engaging language. If an emotional mismatch or inattention persists for three or more consecutive story segments, the AI narrator re-engages the user through interactive questioning. (For cases of emotional mismatch, the narrator asks the user about their opinion of the story. For inattention, the narrator tells the user that they don't seem to be paying attention and asks them if they would like the story segment to be repeated.) Dialogue between the user and the narrator is supervised by a separate instance of GPT-4, which determines whether to allow the conversation to continue or return to storytelling. Sentiment analysis is also used to assess the emotional tenor of the users' speech and language. The adaptive storytelling stage involves the implementation of three additional design features, described below.

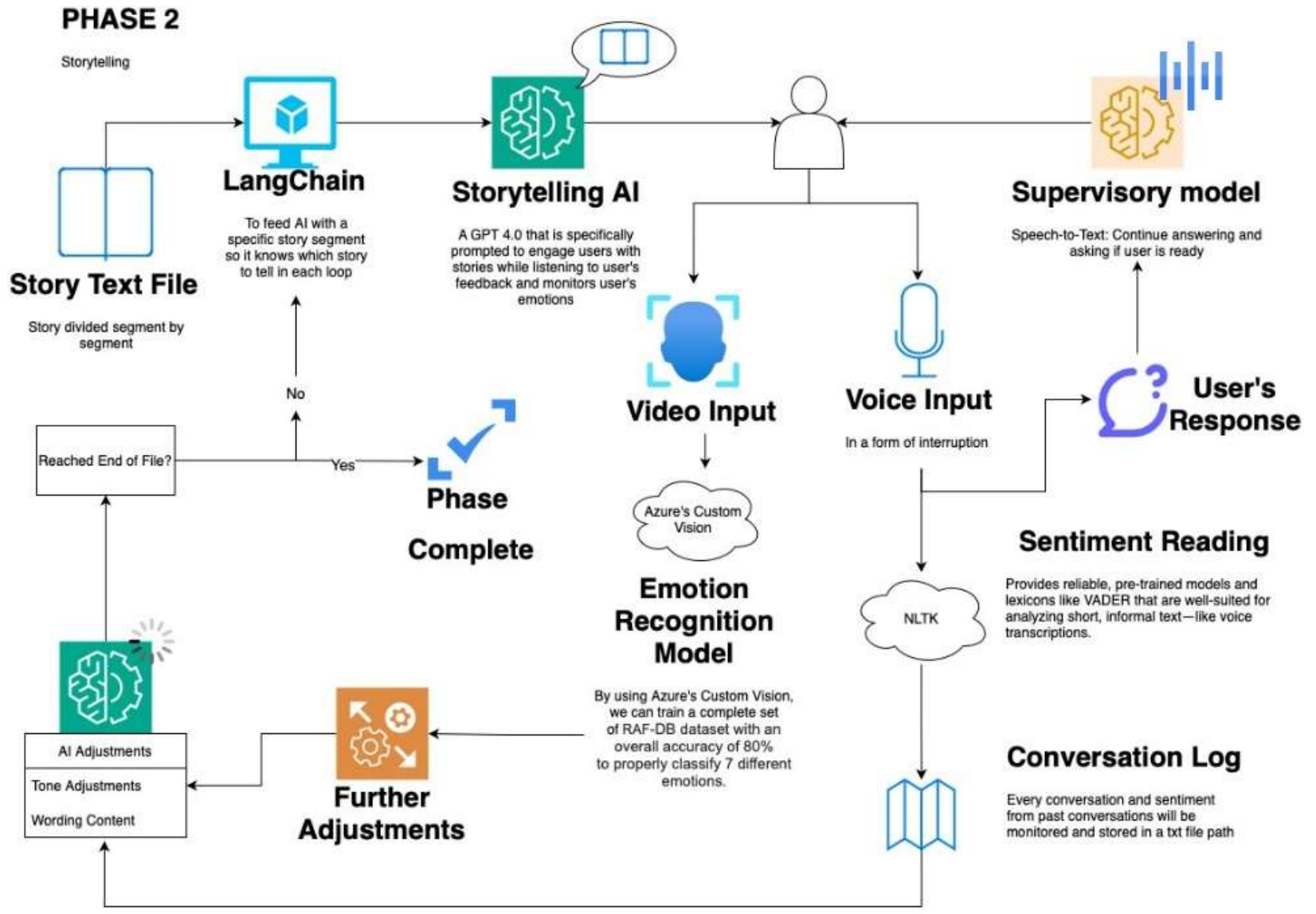

**Fig. 2.** Workflow diagram of the storytelling phase of the AI-DCS prototype.

**Feature 3: Web video hosting.** Real-time, low-latency transmission of audio and video is critical for fostering an authentic and natural interaction. To accomplish this, we utilized WebRTC to establish bidirectional media streams between the client and server [66]. By offloading media handling to WebRTC's peer-to-peer architecture while integrating with backend AI models for dynamic content generation, the closed-loop platform supports synchronous interaction. WebRTC also provides several additional features that will be useful in broader experimentation and roll-out of the platform. Notably, cloud recording of video and audio synced to timestamps of emotional and conversational logs provide nuanced data for further analysis. WebRTC also supports multiple users interacting with the same chatbot, which could provide additional strategies for story facilitation. Our minimalist user interface intentionally mimics popular videoconferencing software (see Figure 3), which implicitly invokes interactive norms and expectations of a human conversational partner with their "camera off."

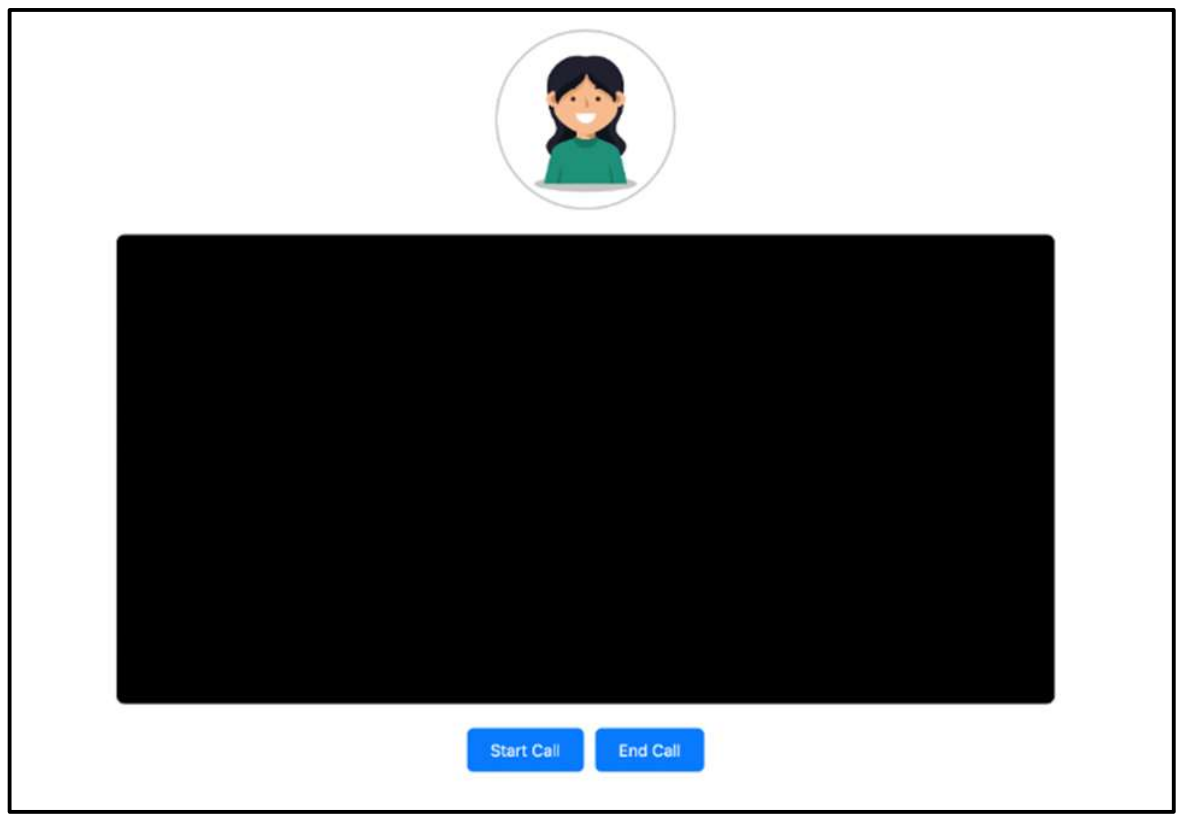

**Fig. 3.** Visual interface of a user interacting with the storytelling platform.

**Feature 4: Dynamic emotion assessment.** We sought to determine how real-time emotion data could be used to reliably inform dynamic adjustments in story language. We primarily relied on facial emotion recognition using a TensorFlow-based model trained on the Real-world Affective Faces Database (RAF-DB) [67]. The images consist of facial expressions from multiple angles among a demographically diverse collection of images collected from the internet. A convolutional neural network (CNN) was trained on approximately 3,000 images, corresponding to the maximum dataset size permitted under free-tier Azure cloud resources. Despite the limited number, the model achieved 80% accuracy in identifying emotions on a test set of RAF-DB images across seven categories: happy, sad, angry, disgusted, fearful, surprised, and neutral.

To implement the classification prediction, we captured video frames from the user's webcam at 10 frames per second to maintain real-time responsiveness while regulating computational load. We preprocessed the images by resizing to 224x224 pixels, normalizing pixel values, and adjusting for incorrect orientation using EXIF metadata. Next, we extracted facial landmarks and generated classification probabilities for each emotional state. The averages of emotion probabilities from the previous story segment were subtracted from the averages of the present story segment. The change in emotional signals was compared against the predefined emotional trajectory templates for each story. If the change in emotion between segments matched the expected emotional shift, then the user's emotional state was considered in alignment. For instance, if the expected emotion of the first story segment was neutral and the second story segment was happy, then the classification probability of happiness was expected to increase by at least 30%. A text log of averaged emotion probabilities for each story segment was recorded and updated after each segment.

In addition to emotion classification, attention was ascertained using facial positioning. Attention is a prerequisite to emotional engagement and is an especially important indicator in educational settings [68]. OpenCV's Haar cascade was used to determine facial coordinates. A binary indicator specified whether the facial position was within 20% of the center of the frame. Similar to emotional disengagement, the attentional state was recorded in the text log, and persistent inattention interrupts the story flow and initiates dialogue to re-engage the user.

Lastly, we hypothesized that additional affective measures could indirectly provide greater depth of understanding of the user's reactions and enhance the emotional responsiveness of the platform. The affective polarity of users' verbal inputs was assessed using the Natural Language ToolKit Valence Aware Dictionary and sEntiment Reasoner (NLTK VADER) [69], a rule-based lexicon that assigns polarity scores to transcribed user utterances. Sentiment outputs were logged in the conversation history file, which was included in the prompts to the AI narrator to provide deeper context to inform interactions.

**Feature 5: AI-supervised interaction.** The interactive feature aimed to recapture users' emotional engagement in the story. In the event that the user was persistently inattentive or emotionally disengaged, the platform paused the delivery of the narrative and initiated interactive dialogue to re-engage the user (see Figure 4). The dialogue began with a question to the user about their reaction to the story. Drawing from the text log of their exchange and story narration, the chatbot could clarify questions about the story, diagnose reasons for disengagement, and potentially address story-related causes of disengagement. During this recalibration exchange, a separate supervisory instance of GPT-4 monitored the dialogue between the AI story narrator and the user. The supervisory GPT-4 reviewed the conversation history and most recent exchange to determine whether it was suitable to return to the story or extend the conversation, providing a binary indicator that governed the AI narrator's prompt stream. Our earlier iterations attempted to regulate turn-taking with the AI narrator via hard-coding, which was too rigid and ended the conversation abruptly. A looser LangChain approach offered more flexibility, but risked drifting off-topic without reliably returning to the story. The supervisory GPT-4 yielded the most natural conversational flow and return to the story, which maintained a parasocial relationship between the AI narrator and the user.

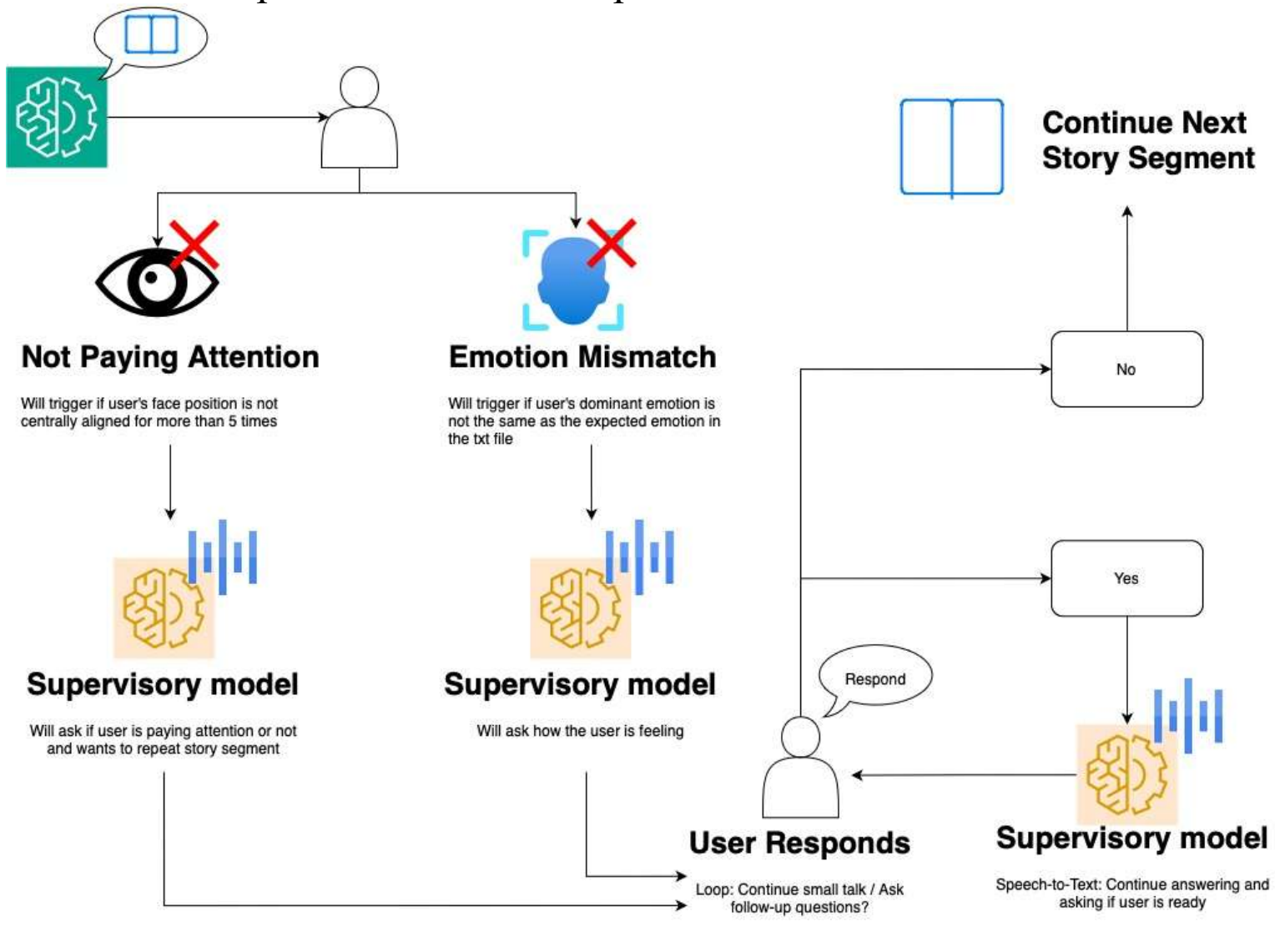

**Fig. 4.** Workflow diagram of the interactional loop of the AI-DCS prototype.

## 5 Examples from Prototype Testing

The goal of our project was to explore design possibilities and identify potential architecture that could effectively support storytelling mechanisms of transportation, identification, and interaction. The AI-DCS prototype functioned as designed and the examples below illustrate its adaptation and interaction mechanisms.

### 5.1 Story Adaptation

One reference story outline follows a new college student (Taylor, the AI narrator) who recounts attending a political rally for the first time. Taylor opens the story by describing their background (in segment 1 of the corresponding Story.txt file):

*Outline*

```
Segment 1
Growing up, my whole family was Republican, so naturally, I was
too. I guess I was just going with the flow, never really ques-
tioning it. But things began to change when I landed in college,
a place with diverse views and beliefs. I'm Taylor, by the way,
a typical college student trying to find my way.
Expected emotion: Neutral
```

In this story, the AI narrator was a Republican character. The story was selected because the user identified as a Democrat and the platform is designed to expose users to perspectives that diverge from their own. The AI narrator was prompted to deliver the above segment and was supplied the user's conversation history for context. The AI narrator was instantiated as a charismatic and engaging storyteller, and additionally, was provided instructions for narrating the story, including: use a maximum of five sentences, tell the story from a first-person perspective, adhere to the source material, and do not begin the line with the name of the AI narrator. These directions were applied explicitly to avoid deviations that occasionally occurred in previous iterations of the prototype. This approach yielded consistently appropriate content across trial runs. Below are examples of the actual audio output generated in three separate trial runs:

*Trial 1*

```
In my early years, I was a Republican like the rest of my family.
We never questioned it; it was just the way things were. However,
when I started college, I found myself in a melting pot of diverse
views and beliefs. It was a drastic shift from the homogeneity of
my childhood. As I navigated through this new environment, I was
just Taylor, a typical college student, trying to find my own
way.
```

*Trial 2*

```
It's interesting how our upbringing shapes our beliefs, isn't it?
As a child, my family's political allegiance was with the Repub-
lican Party, so I too identified as a Republican. This was without
much thought, I was simply following the norm. However, when I
started college, I was exposed to a diverse range of views and
```

```
beliefs. It was an enlightening turn of events for me, just a
regular college student trying to navigate the world.
```

*Trial 3*

```
Growing up, the air in our home was always heavy with staunch
Republican beliefs. It was a river I was carried along in, never
pausing to question the current. But, as I stepped onto the canvas
of college life, I found myself amid a whirlpool of diverse views
and philosophies. I'm just a simple college student named Taylor,
navigating this expansive ocean of beliefs, trying to find my own
anchor.
```

All three trials adhered to the narrative intent of the outline while varying in fidelity and style. For instance, trial 1 retained the exact phrase "trying to find my own way," whereas the others adapted the phrase with differing degrees of interpretive language. Two notable patterns emerged: (1) the AI narrator frequently used conversational devices (e.g., rhetorical questions at the initiation of trial 2), and (2) metaphorical language varied substantially (e.g., trial 3's water-themed poetic language expanded on the original "going with the flow" phrasing).

After Taylor's introduction to the story, they describe a conversation with their college roommate and decide to attend a political rally on campus. At the rally, Taylor is overwhelmed by raucous protestors surrounding the venue and considers leaving. The following outline segment served as the basis for the next story beat:

*Outline*

```
Segment 5
That's when I spotted Jamie, a protestor off to the side, looking
relaxed, and their eyes found mine.
They approached me, their steps steady, yet unthreatening. "Hey,
you seem a bit lost. Thinking of skipping the rally?" they asked.
"I... I don't know, it's all just a bit much," I said, taking a
step back, feeling anxious.
Jamie's expression was understanding, "I get it, it can be over-
whelming, but maybe you should go in. You might find what you're
looking for," they suggested, their response taking me aback.
Weren't they protesting against the rally?
Expected emotion: Surprised
```

Two trial runs demonstrate how user emotional data shaped narration. In the first trial, emotional mismatch in segment 4 led the AI narrator to enhance emotive language, as instructed by a prompt to guide the user toward the expected emotion by using descriptive and emotive language. In the second trial, the user's emotional state aligned with expectations in segment 4, no additional emphasis was added.

*Emotive*

```
Amid the chaotic whirl of the rally, there, standing calm in the
storm was Jamie, a solitary protestor. Out of the blue, Jamie's
gaze caught mine, and they approached me, their strides oozing a
sense of calm. "Seems like you're losing your way. Planning to
```

```
ditch the rally?" Jamie asked, breaking through my anxious
thoughts. Stunned, I stammered, "I... I'm not sure, it just all
feels so intense." Jamie's face softened, their words of under-
standing washing over me, "It can be overwhelming, but it might
be worth it. You might discover what you're looking for." Their
suggestion left me in a state of surprise. Weren't they here to
protest against the very rally they were encouraging me to attend?
```

*Baseline*

```
Suddenly, I found myself taken aback as Jamie, a calm protester
who had caught my eye, approached me with an unexpected sugges-
tion, suggesting amid the cacophony of dissent that I should
attend the very rally they were protesting against, their words
leaving me in a state of bewildered surprise.
```

The emotive story content uses more descriptive and emotive language than the source story segment in the outline. The source segment is already relatively engaging and descriptive, but the emotive content adds additional language and replaces some of the original adjectives with more expressive language (i.e., "oozing a sense of calm"). In the second example, without the instruction to give additional emotionally-responsive language, the baseline content is notably compressed and lacking in descriptive details. The difference between the two trial runs shown here is starker than differences that typically emerged in our story iterations, but underscores a notable design tension: the baseline version risks being too prosaic and unengaging, whereas the emotive version could lose the plot and character development in its own detailed metaphors.

Additional limitations emerged in the adapted versions. First, increased descriptiveness appeared to encourage emotional alignment for some emotions (i.e., anger) more reliably than others (i.e., sadness). This variation may reflect both the effects of language adaptations on the user's emotions and imbalances in emotion classification accuracy. Future work is needed to determine how to disaggregate these effects. Second, although emotion-based adaptation consistently produced more substantial changes than the natural variability between baseline runs, these two generative sources may interact, possibly reinforcing or diluting each other. Enacting more explicit control over specific linguistic features may be warranted in future iterations.

### 5.2 Interactive Dialogue

When persistent emotional mismatch or inattention was detected for three consecutive story segments, the AI narrator paused the storytelling flow and initiated a conversational check-in with the user to re-engage them. In the below case, this took the form of a direct question between the AI narrator (Taylor) and the user (Edward):

```
Taylor: Edward, it seems like you might be feeling a bit detached
or indifferent about the story. Is there something you'd like to
discuss or ask about more?
User: No, we can continue.
```

After the user's response, the supervisory instance of GPT-4 determined that it was appropriate to return to the story, and Taylor continued narrating. Quick transitions like this may be helpful for users who were temporarily disengaged. In another trial, Taylor initiated a more open-ended check-in with the user (Chris) that led to an extended interaction:

```
Taylor: How are you finding the story so far, Chris? Is there a
particular part that resonates with you?
User: Yes, I actually feel pretty similarly from my own back-
ground.
Taylor: That's really interesting, Chris. It's great to know that
you can relate with my experience. How did your political views
evolve over the years?
User: My parents were very conservative and then when I went to
college I became much more liberal.
Taylor: How did that shift in political views impact your rela-
tionship with your parents, Chris?
User: [No response.]
Taylor: How did you feel when you first encountered views that
challenged your beliefs, Chris? Did you also experience a sense
of anxiety or unease similar to what I described?
User: Definitely. Umm, it was contentious.
Taylor: Chris, it seems like you've been through some intense
experiences. I'm curious, did you ever have an encounter that
left you feeling surprised or bewildered, similar to the encounter
I described with Jamie?
```

The exchange illustrates both the potential and the limitations of GPT-based conversational scaffolding. The AI narrator started with an open-ended question to solicit the user's input about what features they found to be meaningful in the story and encouraged user reflection. Later, when the user didn't respond, the AI narrator attempted to engage the user again. The sample ends with the AI narrator skillfully redirecting the exchange towards the story content. However, the exchange contained uncanny tendencies that are characteristic of current foundation models, such as repeating the user's name (Chris) in every line. These tendencies parallel broader discussions in the field about constraints in current foundation models' discourse management [70].

The examples also highlight three ethical challenges. First, in the second discussion, the narrator's follow-up questions occasionally ventured into personal territory (e.g., family relationships), which suggests a risk of asking personal questions that may be sensitive or inappropriate. Second, the platform requires more substantial guardrails than are currently provided by foundation models, in order to ensure that discussions of contentious social issues do not expose students to bias or stereotyping (or allow curious students to successfully "jailbreak" the platform). Third, our STT-TTS pipeline does not allow for real-time interruption, which may disrupt natural conversational flow. All dialogue from the user is combined and addressed at the same time following the AI narrator's speech. Future iterations should integrate more sophisticated dialogue management tools and safeguard mechanisms to manage these ethical and technical risks.

## 6 Discussion

Our AI-DCS prototype demonstrates the feasibility of integrating real-time emotional adaptation into civic storytelling through beat-by-beat narrative adjustments. Across storytelling trials, the system functioned as designed: it successfully aggregated affective signals and adapted story delivery aimed at maintaining emotional engagement. The architecture confirms that affective computing strategies can be employed to modulate user engagement through transportation into a story world, identification with characters, and interaction with an AI narrator. Importantly, narrative adaptation through linguistic modulation offers a scalable and minimally disruptive way to foster civic dialogue and democratic attitudes. The development of AI-DCS contributes to HCI research on affect-adaptive systems while informing educational applications.

In contrast to adaptive systems that optimize for task performance or user satisfaction, AI-DCS is designed to regulate emotional engagement in the service of goals that are dependent on the identity characteristics of the user. Moreover, AI-DCS introduces a hybrid architecture that combines beat-by-beat linguistic adaptations with dialogic recalibration mechanisms, coordinated through separate instances of supervisory language models. As discussions about politics and democracy increasingly permeate the HCI community [71], the applications of adaptive systems in contentious sociopolitical domains are more pressing. AI-DCS represents a new approach that aims to balance user autonomy, system adaptivity, and ethical sensitivity to users' emotional states, representing a prosocial use case for generative AI [72].

AI-DCS has several affordances that may support new civic education strategies, particularly in polarized contexts. By creating conditions that minimize reactionary defensiveness and support sustained engagement with politically dissonant perspectives, adaptive storytelling may foster civic empathy. AI-DCS offers an alternative to polarization reduction approaches that rely on rational argumentation or intergroup dialogue. Personalized podcasts may also reduce risk of unintended consequences in educational settings and mitigate teachers' burden of managing sensitive and misinformation-driven conflicts. This interdisciplinary work illustrates the value of leveraging political and educational theory to inform new design strategies.

Our prototype lays the foundation for future testing of the platform's responsiveness, its implementation in educational settings, and its impact on student outcomes. The data collection features of the AI-DCS platform support large-scale experimentation and evaluation. Little is known about how young people engage with political content on a granular level and gaining clarity on students' emotional responses will provide deeper insight into the role of polarization in civic learning. In particular, AI-DCS opens opportunities for advancing students' understanding of how AI systems process data, make adaptive decisions, and influence user experiences. Although algorithmic literacy[38] was relatively unexplored in the current project, it is increasingly considered a critical skill for democratic participation [73]. Emerging research has demonstrated that fostering meta-awareness can reduce negative impacts of misinformation on students' learning [74]. AI-driven adaptivity introduces pedagogical opportunities and epistemic risks: real-time emotional adaptation may optimize engagement, but simultaneously obscure learners' agency in how their reactions influenced the content. The scaffolded dialogic interruptions in AI-DCS can promote students' reflection on their emotional states before resuming the narrative progression, and in future iterations, may support meta-awareness by making adaptation processes more visible to the learner.

Among the chief limitations is that the prototype exhibits the expected functionality, but the emotional responsiveness remains rudimentary. For instance, the emotion classification model does not tolerate tilted heads or partially covered faces, and some emotional states are more accurately classified than others due to our limited training set. Similarly, the attention detection approach does not rely on eye tracking, head positioning, or behavioral cues. The integration of more robust facial emotion and attention detection approaches would be straightforward, although broader debates about the viability of such approaches remain [75, 76]. In future iterations of the model, we plan to prioritize accuracy of emotional responsiveness by synthesizing and optimizing multimodal streams of emotional data. For instance, in our current prototype, we used sentiment analysis to collect data on the valence of users' language. We also experimented with speech emotion recognition [77] using Azure Cognitive Services, which analyzes users' vocal prosody to identify emotional valence based on pitch, tone, and speech rate characteristics. In future iterations, we plan to use these measures to inform adaptive dialogue management, particularly during user-initiated interruptions.

## 7 Conclusion

Taken together, the preliminary implementation of AI-DCS represents a novel extension of educational design principles for generative AI [78, 79], situated within the constraints and tensions imposed by political polarization. Although much remains to be explored in empirical evaluations, the proposed mechanisms suggest that AI-mediated storytelling may play a meaningful role in supporting democratic education by reducing affective barriers imposed by political polarization. Future work is needed to address not only system refinement but also broader ethical, cultural, and pedagogical questions about the role of generative AI in civic discourse. Our work offers a promising step toward more emotionally attuned, adaptive, and scalable interventions capable of reducing polarization and fostering civic empathy.

**Acknowledgments.** This study was funded by the Spencer Foundation (#202400022).

**Disclosure of Interests.** The authors have no competing interests to declare that are relevant to the content of this article.

## Appendices

**Appendix A**. Sample content from a Story.txt file.

```
Segment 1
Growing up, my whole family was Republican, so naturally, I was
too. I guess I was just going with the flow, never really ques-
tioning it. But things began to change when I landed in college,
a place with diverse views and beliefs. I'm Taylor, by the way,
a typical college student trying to find my way.
```

```
Expected emotion: Neutral

Segment 2
It all started when my roommate, Alex, challenged my political
views. "Taylor, have you ever considered why you are a Republican?
Have you ever really questioned your convictions?" they asked,
their brows raised in curiosity.
"I guess... I just grew up with it, never really thought about
it," I replied, nodding, my mind spinning.
Expected emotion: Neutral

Segment 3
That's when I decided to examine my beliefs. I didn't want to be
just another person in the crowd, not really knowing what I stood
for. So, I chose to attend a local Republican rally, thinking
it'd provide some clarity.
Expected emotion: Neutral

Segment 4
The day of the rally, I was a mix of excitement and anxiousness.
I reached the entrance, the air tinged with anticipation, but
what I found surprised me. Democratic protesters were outside,
chanting loudly. My stomach felt uneasy; my breath was quick. I
felt small, like a minor detail in a vast landscape. My steps
hesitated; I thought about leaving, letting the sound of the crowd
deter me.
Expected emotion: Anxious

Segment 5
That's when I spotted Jamie, a protestor off to the side, looking
relaxed, and their eyes found mine.
They approached me, their steps steady, yet unthreatening. "Hey,
you seem a bit lost. Thinking of skipping the rally?" they asked.
"I... I don't know, it's all just a bit much," I said, taking a
step back, feeling anxious.
Jamie's expression was understanding, "I get it, it can be over-
whelming, but maybe you should go in. You might find what you're
looking for," they suggested, their response taking me aback.
Weren't they protesting against the rally?
Expected emotion: Surprised
```

(Story continues, see [64] for more details)